\documentclass[preprint,12pt]{elsarticle}
\usepackage{amssymb,amsfonts,amsthm}
\usepackage[fleqn]{amsmath}
\usepackage{geometry}
\usepackage{graphics,subfigure}
\usepackage{hyperref}
\usepackage{pifont}
\usepackage{natbib}
\usepackage{color}
\usepackage{bbm}
\usepackage{bm}
\usepackage{appendix}
\numberwithin{equation}{section}
\journal{Proceedings of the Royal Society A}

\begin{document}

\begin{frontmatter}
\title{Influence of the growth gradient on surface wrinkling and pattern transition in growing tubular tissues}

\author[add1]{Rui-Cheng Liu}
\author[add1,add2]{Yang Liu\corref{cor1}}
\ead{tracy\_liu@cityu.edu.hk}
\author[add1,add2]{Zongxi Cai}
\cortext[cor1]{corresponding author}
\address[add1]{Department of Mechanics, School of Mechanical Engineering, Tianjin University, Tianjin 300350, China}
\address[add2]{Tianjin Key Laboratory of Modern Engineering Mechanics, Tianjin 300350, China}

\begin{abstract}
Growth-induced pattern formations in curved film-substrate structures have attracted extensive attentions recently. In most existing literature, the growth tensor is assumed to be homogeneous or piecewise homogeneous. In this paper, we aim at clarifying the influence of a growth gradient on pattern formation and pattern evolution in bilayered tubular tissues under plane-strain deformation. In the framework of finite elasticity, a bifurcation condition is derived for a general material model and a generic growth function. Then we suppose that both layers are composed of neo-Hookean materials. In particular, the growth function is assumed to decay linearly either from the inner surface or from the outer surface. It is found that a gradient in the growth has a weak effect on the critical state, compared to the homogeneous growth type where both layers share the same growth factor. Furthermore, a finite element model is built to validate the theoretical model and to investigate the post-buckling behaviors. It is found that the associated pattern transition is not controlled by the growth gradient but by the ratio of the shear modulus between two layers. Different morphologies can occur when the modulus ratio is varied. The current analysis could provide useful insight into the influence of a growth gradient on surface instabilities and suggests that a homogeneous growth field may provide a good approximation on interpreting complicated morphological formations in multiple systems.
\end{abstract}
\begin{keyword}
Surface instability \sep Graded growth  \sep Non-linear elasticity  \sep Finite element analysis \sep Pattern transition 
\end{keyword}
\end{frontmatter}

\section{Introduction}
Ubiquitous surface patterns, ranging from a microscopic scale to a macroscopic one, support the foundation of regular biological functions in biological systems \cite{1}. Specifically, in mammalian brains \cite{2,3,4}, fruits and vegetables \cite{5,6}, various growth-induced surface morphologies can be observed at different growth periods. In general, brain surface is smooth at the new born stage and becomes wrinkled due to differential growth between the gray matter and the white matter \cite{2}. On the mechanical side, such phenomenon is in accordance with structure instability in classical mechanics. On the other hand, soft tissues can be modeled by incompressible hyperelastic materials and the material properties can regulate the final morphology. Therefore, it is of significant importance to reproduce the deformation process according to some growth models. In some existing studies, the growth model proposed by Rodriguez et al. \cite{7} was employed, in which the deformation gradient is decomposed by the multiplier of an elastic deformation gradient and a growth tensor. Based on such a theory, Ben Amar and Goriely \cite{8} established a finite growth theory for soft tissues and studied several fundamental examples. Recent advances on growth-induced instabilities can be found in the review articles \cite{1,9} and the monograph by Goriely \cite{10}.

This study is concerned with tubular tissues, such as esophagi and intestines, which are widely distributed in animal bodies. In practice, these tissues are mainly composed of mucosa and submucosa \cite{11}. Generally speaking, the final pattern formed on the surface of tubular tissues is dependent on various factors, including the initial material properties \cite{12,13,14}, the growth pathways \cite{15,16}, and the remodeling during growth process \cite{17,18,19}, etc. In spite of these facts, a bilayered model within the continuum mechanical framework paves a convenient way to unravel the mechanism behind various patterns using analytical or numerical approaches \cite{12,13,14,15,16,17}. It should be pointed out that, the growth tensor is diagonal and homogeneous in most existing investigations. In addition, the tissue is modeled by isotropic and homogeneous material. In doing so, residual stresses and pattern formations can be induced by certain geometric restrictions on the outer boundary \cite{11,14} or differential growth between inner and outer layers \cite{12,13,15,16,17}. In our previous studies, the scaling laws of the critical growth factor and the critical wavenumber and the amplitude equation for wrinkling mode were derived by virtue of the WKB technique and the virtual work method, respectively \cite{20,21}. Therefore, the current investigation can be viewed as a series work.

It is worth mentioning that, in practical systems and realistic growth processes, either the material property or the growth factor can be inhomogeneous \cite{22,23}. For instance, prestrain in arteries is supposed to be induced by heterogeneous growth \cite{24}. Furthermore, plants can adjust their growth and produce complex behaviors according to various environmental stimuli \cite{25}, and the final bacterial biofilm morphology can be altered by nonuniform growth \cite{26}. On the other hand, certain boundary constrains also produce inhomogeneous growth. Then B\"ol and Albero \cite{27} proposed a theoretical model for describing the effect of inhomogeneous growth in elastic bodies and modeled several constraint growth cases. Recently, Lee et al. \cite{28} defined a geometric incompatibility tensor and implemented it into nonlinear finite element framework in order to explore inhomogeneous growth-induced deformation and pattern formations. As illustrative examples, brain atrophy, skin expansion, and cortical folding were analyzed numerically based on their model. In our previous study \cite{29}, a thorough analysis was performed to reveal the effect of modulus gradient on the surface instabilities of growing tubular tissues. It turns out that modulus gradient can alter pattern evolution. Bearing in mind that much existing literature exploits the assumption of homogenous growth, it is well-motivated to elucidate the influence of the growth inhomogeneity on surface instabilities in growing tubular tissues. This is also the aim of the sequel study.

The rest of this paper is organized as follows. In Section 2, we identify the deformation prior to surface instability and present the incremental theory for a general material model and a general growth gradient under the plane-strain setting. A linear bifurcation analysis is carried out in Section 3. A finite element model for graded growth is established and then validated in Section 4. We exploit the finite element model to trace the post-buckling evolutions and illustrate two distinct patterns in Section 5. Finally, we give some concluding remarks in Section 6.

\section{Theoretical model}
\begin{figure}[!h]
	\centering\includegraphics[scale=0.45]{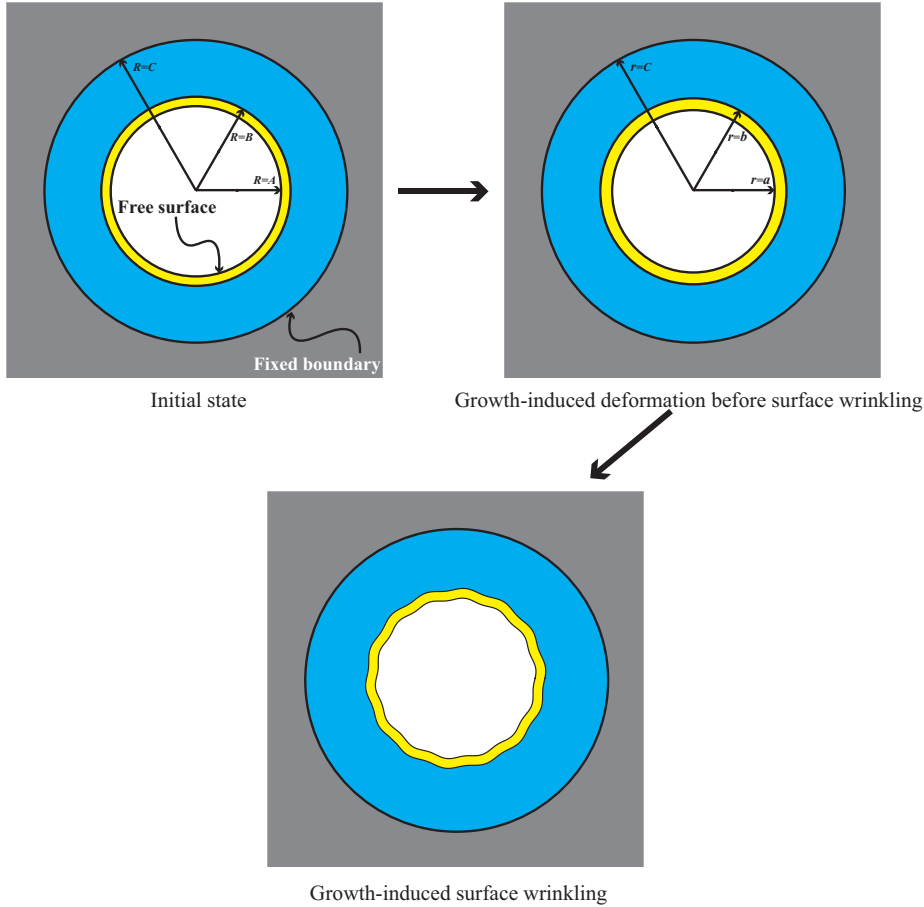}
	\caption{Two-dimensional sketches of the initial state, the basic state, and the wrinkled state of a bilayered tubular tissue.}\label{fig1}
\end{figure}

Consider a growing tubular tissue consisting of two layers. As shown in Figure \ref{fig1}, in the initial state $\mathcal{B}_0$, the inner and outer radii are given by $A$ and $C$, respectively, and the interfacial radius is denoted by $B$. It is assumed that the inner surface is traction-free while the outer surface is attached to a rigid body so that the outer radius is fixed in the growth process. Meanwhile, the interface keeps perfectly bonded during the deformation, so both the traction and displacement are continuous across the interface. Suffering a graded growth effect, this tubular tissue will grow thicker to form the basic state $\mathcal{B}_r$, and the inner and interfacial radii develop into $a$ and $b$, respectively. Note that in this study we only focus on a plane-strain deformation and then the axial dimension will not be mentioned throughout the analysis. When the growth factor passes a critical value, a wrinkled pattern will emerge in the circumferential direction, and the bifurcated state is called $\mathcal{B}_t$. In this section, we shall analytically characterize the basic state $\mathcal{B}_r$ and derive the incremental equation from $\mathcal{B}_r\longrightarrow\mathcal{B}_t$ for a general material model and a general growth function.

Since there are two layers, we first define the notation convention. In the subsequent analysis, a quantity with a bar belongs to the outer layer, or otherwise it is owned by the inner layer. For instance, the strain-energy functions for the inner and outer layers are expressed as $W$ and $\bar{W}$, respectively. Moreover, the bar will be dropped if a notation is applicable to both layers. For example, the growth function is expressed by $g(R)$ and it is evaluated in the inner layer when $A<R<B$ or computed in the outer layer otherwise. We emphasize that the theoretical derivations for both layers are quite similar. Therefore, only the details for the inner layer will be written out, and the counterparts for the outer layer can be obtained by proper variable substitutions.

We adopt the cylindrical polar coordinate system for convenience. For a plane-strain problem, the coordinates for a material point in $\mathcal{B}_0$ and $\mathcal{B}_r$ are represented by $(R,\Theta)$ and $(r,\theta)$, respectively. Meanwhile, the corresponding position vectors are given by $\bm X$ and $\bm x$, and the common orthonormal basis \{$\bm e_r, \bm e_\theta$\} is used. In doing so, the deformation gradient for the inner layer associated with $\mathcal{B}_0\longrightarrow\mathcal{B}_r$ is given by $\textbf{F}=\partial {\bm x}/\partial {\bm X}$. Considering that the primary deformation is axisymmetric and excludes torsion, we write the component form of $\textbf{F}$ by
\begin{equation}
	\textbf{F}=\left[\begin{array}{cc} \dfrac{\textrm{d} r}{\textrm{d} R} & 0 \\ 0 & \dfrac{r}{R}\end{array}\right].
	\label{eq2_1}
\end{equation}
For convenience, we use $\lambda_1=\textrm{d} r/\textrm{d} R$ and $\lambda_2=r/R$ to denote the principal stretches. Note that 1 and 2 correspond to $r$- and $\theta$-directions, respectively.

To take the growth effect into consideration, we refer to the multiplicative decomposition theory \cite{7} and hence the deformation gradient is split as
\begin{equation}
	\textbf{F}=\textbf{A}\textbf{G},\label{eq2_2}
\end{equation}
where $\textbf{G}$ expresses the growth tensor and $\textbf{A}$ stands for the elastic deformation gradient tensor. In particular, we suppose that they are given by the following formulas:
\begin{equation}
\textbf{G}=\left[\begin{array}{cc}  g_1(R) & 0 \\ 0 & g_2(R)\end{array}\right],
~~\textbf{A}=\left[\begin{array}{cc}  \alpha_1 & 0 \\ 0 & \alpha_2\end{array}\right],\label{eq2_3}
\end{equation}
where the growth factor $g_i(R)$ $(i=1,2)$ is dependent on the radial position in a graded growth situation and $\alpha_i$ $(i=1,2)$ corresponds to the radius-dependent elastic principal stretch. 

In general, soft tissues can be regarded as incompressible hyperelastic materials, so we denote the strain-energy function for the inner layer by $W(\mathbf{A})=W(\alpha_1,\alpha_2)$. In this way, the Cauchy stress tensor $\bm \sigma$ reads
%\begin{equation}
%	\textbf{S}=J\textbf{G}^{-1}\left(\frac{\partial W}{\partial \textbf{A}}-p\textbf{A}^{-1}\right),\label{eq2_4}
%\end{equation}
\begin{equation}
\bm{\sigma}=\textbf{A}\frac{\partial W}{\partial \textbf{A}}-p\textbf{I},\label{eq2_5}
\end{equation}
where $\textbf{I}$ denotes the second-order identity tensor and $p$ is the Lagrange multiplier, or hydrostatic pressure, enforcing the incompressibility condition. 
% and $p$ is the Lagrange multiplier, or hydrostatic pressure, enforcing the incompressibility condition. By means of the Nanson's formula, the  is given by

The components of Cauchy stress are given by 
\begin{align}
\sigma_{ii}=\sigma_i-p,~~\sigma_i=\alpha_iW_{,i},~~\mbox{no~summation~on}~i.\label{eq2_6}
\end{align}
We define in this study that a comma behind a quantity indicates differentiation with respect to the corresponding variable, such as $W_{,2}=\partial W/\partial \alpha_2$.

Neglecting the body force, we arrive at the equilibrium equations in the Euler description:
\begin{equation}
	\operatorname{div}\bm \sigma=\textbf{0},\label{eq2_7}
\end{equation}
where the ``div'' means that the operator is evaluated in the current configuration $\mathcal{B}_r$. We point out that the primary deformation is axisymmetric, which results in the only equilibrium equation:
\begin{align}
	\dfrac{\textrm{d}\sigma_{rr}}{\textrm{d} r}+\dfrac{\sigma_{rr}-\sigma_{\theta\theta}}{r}=0.\label{eq2_8}
\end{align}
The traction-free condition on the inner surface can be expressed as
\begin{align}
\sigma_{rr}=0,~~\mbox{on}~~r=a.\label{eq2_9}
\end{align}
In addition, the continuity of traction on the interface supplies
\begin{align}
\sigma_{rr}=\bar{\sigma}_{rr},~~\mbox{on}~~r=b.\label{eq2_10}
\end{align}

Then we substitute (\ref{eq2_6}) into (\ref{eq2_8}) and apply the boundary condition (\ref{eq2_9}) to obtain
\begin{align}
\sigma_{rr}=\int^r_a\dfrac{\alpha_2W_{,2}-\alpha_1W_{,1}}{r}\textrm{d}r.\label{eq2_11}
\end{align}
In addition, the Lagrange multipliers can be determined by 
\begin{align}
	&p=\alpha_{1}W_{,1}+\int^r_{a}\dfrac{1}{r}(\alpha_{1}W_{,1}-\alpha_{2}W_{,2})\textrm{d}r,\notag\\ 
	&\bar{p}=\bar{\alpha}_{1}\bar{W}_{,1}+\int^r_{b}\dfrac{1}{r}(\bar{\alpha}_{1}\bar{W}_{,1}-\bar{\alpha}_{2}\bar{W}_{,2})\textrm{d}r+\int^b_{a}\dfrac{1}{r}(\alpha_{1}W_{,1}-\alpha_{2}W_{,2})\textrm{d}r.
	\label{eq2_12}
\end{align}

It can be deduced from the incompressibility condition $\operatorname{det}\mathbf{F}=\operatorname{det}\mathbf{G}$  and the fixed boundary condition on the outer surface that the deformed radius for any material point is given by  
\begin{align}
r^2=2\int_C^Rg_1(R)g_2(R)R\textrm{d}R+C^2.\label{eq2_13}
\end{align}

Currently, the basic state $\mathcal{B}_r$ has been completely determined. Once the exact forms of $g_1(R)$ and $g_2(R)$ are prescribed, one could obtain all stresses and displacements from (\ref{eq2_11})--(\ref{eq2_13}). 

Next, we intend to derive the linearized incremental equation for further bifurcation analysis. To this end, we signify associations from $\mathcal{B}_0$ to $\mathcal{B}_t$ by a tilde, for instance, the corresponding deformation gradient and nominal stress for the inner layer write $\widetilde{\textbf{F}}$ and  $\widetilde{\textbf{S}}$, respectively. Furthermore, we suppose that the wrinkled state $\mathcal{B}_t$ is attained by superimposing on $\mathcal{B}_r$ an incremental displacement field $\delta \bm x$:
\begin{equation}
	\delta \bm x=u(r,\theta)\bm e_r+v(r,\theta)\bm e_\theta,
\end{equation}
where $u(r,\theta)$ is the  increment in the radial direction and $v(r,\theta)$ the counterpart in the hoop direction.
It can be readily checked that $\widetilde{\mathbf{F}}=(\mathbf{I}+\bm\eta)\mathbf{F}$ where $\bm\eta$ is given by
\begin{equation}
	\bm\eta=\left[\begin{array}{cc} u_{,r} & \frac{1}{r}\left(u_{,\theta}-v\right) \\ v_{,r} & \frac{1}{r}\left(v_{,\theta}+u\right)\end{array}\right].\label{eq2_15}
\end{equation}

To formulate the incremental equation, we define an incremental stress tensor $\bm\chi$ as \cite{30}
\begin{equation}
	\bm\chi^\textrm{T}=J^{-1}\textbf{F}(\widetilde{\textbf{S}}-\textbf{S}).
\end{equation}
In the above expression, we have denoted $J=\operatorname{det}(\textbf{G})$ which signifies the volume increase due to growth, the $\mathbf{S}$ stands for the nominal stress tensor from $\mathcal{B}_0$ and $\mathcal{B}_r$ whose expression can be deduced according to the Nanson's formula and equation (\ref{eq2_5}), and the superscript ``T'' represents transpose. In doing so, the incremental equilibrium equation reads $\operatorname{div}\bm\chi^\textrm{T}=\bm 0$, which further yields
\begin{equation}
	\dfrac{\partial \chi_{rr}}{\partial r}+\dfrac{1}{r}\dfrac{\partial \chi_{r\theta}}{\partial \theta}+\dfrac{\chi_{rr}-\chi_{\theta\theta}}{r}=0,~~~~
	\dfrac{\partial  \chi_{\theta r}}{\partial r}+\dfrac{1}{r}\dfrac{\partial\chi_{\theta\theta}}{\partial \theta}+\dfrac{\chi_{\theta r}+\chi_{r\theta}}{r}=0.\label{eq2_18}
\end{equation}

The linearized incremental incompressibility condition is equivalent to the vanish of the trace of $\bm\eta$ in (\ref{eq2_15}) and is given by
\begin{equation}
	u_{,r}+\dfrac{1}{r}\left(v_{,\theta}+u\right)=0.
	\label{eq2_19}
\end{equation}
In addition, taking the Taylor expansion of $\bm\chi$ in $\textbf{F}$ and neglecting all nonlinear terms furnish 
\begin{equation}
	\chi_{ij}=\mathcal{A}_{jilk}\eta_{kl}+p\eta_{ji}-p^{*}\delta_{ji},\label{eq2_20}
\end{equation}
where $p$ has been defined in (\ref{eq2_5}) and $p^{*}$ stands for the corresponding incremental counterpart. Assuming that the growth tensor $\mathbf{G}$ is independent of the deformation gradient $\mathbf{F}$ and referring to \cite{31,32}, all non-trivial instantaneous moduli $\mathcal{A}_{jilk}$ can be computed according to
\begin{align*}
	&\mathcal{A}_{iijj}=\mathcal{A}_{jjii}=\alpha_i\alpha_jW_{,ij},~~\textrm{no~summation~on~}i~\textrm{or}~j,\\
	&\mathcal{A}_{ijij}=\dfrac{\alpha_iW_{,i}-\alpha_jW_{,j}}{\alpha_i^2-\alpha_j^2}\alpha_i^2,~~\alpha_i\neq \alpha_j,~~\textrm{no~summation~on~}i~\textrm{or}~j,\\
	&\mathcal{A}_{ijji}=\mathcal{A}_{ijij}-\alpha_iW_{,i},~~i\neq j,~~\textrm{no~summation~on~}i~\textrm{or}~j.
\end{align*}

The incremental boundary conditions write
\begin{align}
	&\bm{\chi m}\big|_{r=a}=\bm 0,\notag\\
	&\bar{u}\big|_{r=C}=\bar{v}\big|_{r=C}=0,
	\label{eq2_21}
\end{align}
where $\bm m=-\bm e_r$ corresponds to the outward unit normal vector to the inner surface. On the other hand, the continuity conditions on the interface can be expressed by
\begin{align}
&(\bm{\chi}-\bm{\bar{\chi})m}\big|_{r=b}=\bm 0,~~(u-\bar{u})\big|_{r=b}=0,~~(v-\bar{v})\big|_{r=b}=0.\label{eq2_22}
\end{align}
In the light of the stress components in (\ref{eq2_20}), we can write explicitly the incremental equations, the boundary conditions, and the continuity conditions. To save length, these details are illustrated in Appendix A.

In summary, the theoretical model for buckling analysis is established in the framework of nonlinear elasticity for a bilayered tubular tissue subjected to a graded growth and to a plane-strain deformation. In particular, the exact forms of the strain-energy function and the growth function are not assigned. Later, a bifurcation analysis will be conducted to determine the bifurcation threshold of surface wrinkling.

\section{Bifurcation analysis}

In this section, we aim at carrying out a bifurcation analysis based on the linearized incremental equation presented in the previous section. It should be pointed out that we are concerned with the case that a wrinkled profile shall take place in the hoop direction at a critical state. In doing so, we specify the incremental perturbations by the following forms
\begin{align}
	&u=U(r)\operatorname{cos}(n\theta),~~v=V(r)\textrm{sin}(n\theta),~~p^{*}=P(r)\operatorname{cos}(n\theta),\notag\\
	&\bar{u}=\bar{U}(r)\operatorname{cos}(n\theta),~~\bar{v}=\bar{V}(r)\operatorname{sin}(n\theta),~~\bar{p}^{*}=\bar{P}(r)\operatorname{cos}(n\theta),
	\label{eq3_1}
\end{align}
where $n$ is the circumferential wavenumber. It turns out that the unknowns $V(r)$ and $P(r)$ can be expressed in terms of $U(r)$ in accordance with (\ref{eq2_19}) and (\ref{eq2_24}). Then we eliminate $V(r)$ and $P(r)$ in equation (\ref{eq2_23}) to arrive at a fourth-order ordinary differential equation ($\textbf{ODE}$) of $U(r) $. Apply the same procedure to the outer layer, a parallel equation of $\bar{U}(r)$ can be found. To facilitate analysis, we define that a prime implies $\textrm{d}/\textrm{d}r$ and introduce two vector fields $\bm y=[U, U', U'', U''']^\textrm{T}$ and $\bar{\bm y}=[\bar{U}, \bar{U}', \bar{U}'', \bar{U}''']^\textrm{T}$ and rewrite the incremental equations, the boundary conditions and the continuity conditions as follows
\begin{align}
	&\dfrac{\textrm{d}\bm y}{\textrm{d}r}=\mathbf{M}(r,n)\bm y,~~a<r<b,\label{eq3_2}\\
	&\dfrac{\textrm{d}\bar{\bm y}}{\textrm{d}r}=\bar{\mathbf{M}}(r,n)\bar{\bm y},~~b<r<C,\label{eq3_3}\\
	&\mathbf{T}_1(r,n)\bm y=\bm 0,~~\mbox{on}~r=a,\notag\\
	&\mathbf{T}_{2}(r,n)\bar{\bm y}=\bm 0,~~\mbox{on}~r=C,\notag\\
	&\mathbf{T}_3(r,n)\bm y-\bar{\mathbf{T}}_{3}(r,n)\bar{\bm y}=\bm 0,~~\mbox{on}~r=b,\label{eq3_4}
\end{align}
where the coefficient matrices $\mathbf{M}$, $\mathbf{T}_1$, $\mathbf{T}_2$, and $\mathbf{T}_3$ are specified by
\begin{align*}
\mathbf{M}(r,n)=\left[\begin{array}{cccc}0 & 1 & 0 & 0 \\0 & 0 & 1 & 0 \\0 & 0 & 0 & 1 \\ m_{41} & m_{42} & m_{43} & m_{44}\end{array}\right],
\end{align*}
\begin{align*}
\mathbf{T}_1(r,n)=\left[\begin{array}{cccc} \dfrac{1-n^2}{r}\mathcal{A}_{1212} & -\mathcal{A}_{1212} & -r\mathcal{A}_{1212} & 0 \\t_{21} & t_{22} & t_{23} & -r^2\mathcal{A}_{1212}\end{array}\right],~~\mathbf{T}_2(r,n)=\left[\begin{array}{cccc}1 & 0 & 0 & 0 \\0 & 1 & 0 & 0\end{array}\right],\\
\end{align*}
\begin{align*}
\mathbf{T}_3(r,n)=\left[\begin{array}{cccc} \dfrac{1-n^2}{r}\mathcal{A}_{1212} & -\mathcal{A}_{1212} & -r\mathcal{A}_{1212} & 0 \\t_{21} & t_{22} & t_{23} & -r^2\mathcal{A}_{1212}\\ 1& 0 & 0 & 0 \\0 & 1 & 0 & 0\end{array}\right],
\end{align*}
and the elements of $\bar{\mathbf{M}}$ and $\bar{\mathbf{T}}_1$ can be acquired by proper variable substitutions and then are omitted for the sake of conciseness. Correspondingly, the components $m_{ij}$ and $t_{ij}$ of these matrices are shown in Appendix B.

The bifurcation threshold where circumferential wrinkling initiates can be identified by solving the eigenvalue problem of the above linear system, which contains position-dependent coefficients. Once the material models for the inner and outer layers and the growth gradient are specified, one could apply the determinant method \cite{33,34} or the compound matrix method \cite{35,36} to solve the eigenvalue problem numerically in Mathematica \cite{37}. For the sake of definiteness, we now suppose that both the inner and outer layers are composed of incompressible neo-Hookean materials so that 
\begin{align}
W=\dfrac{1}{2}\mu(\alpha_1^2+\alpha_2^2-2),~~\bar{W}=\dfrac{1}{2}\bar{\mu}(\bar{\alpha}_1^2+\bar{\alpha}_2^2-2),\label{eq3_5}
\end{align}
where $\mu$ and $\bar{\mu}$ denote the ground state shear moduli. 

Substituting (\ref{eq3_5}) into (\ref{eq3_2}) furnishes the following fourth-order ordinary differential equation:
\begin{align}
U^{(4)}+\dfrac{2}{r}U'''-\left(\dfrac{3+n^2}{r^2}+\dfrac{n^2r^2}{R^4g_2^4}\right)U''+\left(\dfrac{3+n^2}{r^3}-\dfrac{3n^2r}{R^4g_2^4}\right)U'+(n^2-1)\left(\dfrac{3}{r^4}+\dfrac{n^2}{R^4g_2^4}\right)U=0,\label{eq3_6}
\end{align}
Accordingly, the incremental boundary condition $(\ref{eq3_4})_1$ yields
\begin{equation}
U'''+\dfrac{2}{r}U''-\left(\dfrac{1+n^2}{r^2}+\dfrac{n^2r^2}{R^4g_2^4}+\dfrac{pn^2}{\mu R^2g_2^2}\right)U'+\dfrac{1-n^2}{r^3}U=0,~~~\textrm{on}~~r=a,\label{eq3_7}
\end{equation}
\begin{equation}
-rU''-U'+\dfrac{1-n^2}{r}U=0,~~~\textrm{on}~~r=a.\label{eq3_8}
\end{equation}
In fact, the $R$ can be denoted in terms of $r$ according to (\ref{eq2_13}) and so $g_2$ can also be rewritten as a function of $r$. Note that we have omitted the argument ``$r$'' in (\ref{eq3_6})--(\ref{eq3_8}) for brevity. 

Next, we shall prescribe a detailed formula for the growth gradient. Note that Tallinen et al. \cite{3} reproduced brain morphology by employing an exponentially decayed growth function, and Lee et al. \cite{28} adopted a linearly position-dependent function to model inhomogeneous growth. It is emphasized that the exact growth distribution is complicated and may be affected by many factors. Bearing in mind that the coupling effect between growth and stress (or strain) is neglected, we focus on a special case that the growth functions occupy the same expression in the radial and hoop directions. In addition, we employ two specific cases: The growth function decays linearly from the inner surface and the outer surface is free of growth and the reverse situation. In doing so, we write 
\begin{align}
&\textrm{graded~growth~type~I~(GGI):}~~g_1(R)=g_2(R):=f(R)=(g-1)\left(1-\dfrac{R}{C}\right)+1,\label{eq3_9}\\
&\textrm{graded~growth~type~II~(GGII):}~~g_1(R)=g_2(R):=\widehat{f}(R)=(g-1)\left(\dfrac{R}{A}-1\right)+1,\label{eq3_10}
\end{align}
where $g\geqslant1$ denotes the growth factor that drives the growing process. Actually, the linear distributions in (\ref{eq3_9}) and (\ref{eq3_10}) can be regarded as first-order Taylor expansions for other complex distributions, and it is anticipated that such simple relations could provide some useful insight.

To facilitate the subsequent analysis, we apply the non-dimensional procedure by scaling all quantities of length dimension by $C$ and all quantities of stress dimension by $\bar{\mu}$. In this way, all equations, boundary conditions and continuity conditions are reformulated and we introduce some dimensionless parameters as follows
\begin{align}
\beta=\dfrac{\mu}{\bar{\mu}},~A_*=\dfrac{A}{C},~B_*=\dfrac{B}{C},~a_*=\dfrac{a}{C},~b_*=\dfrac{b}{C}.
\end{align}
When the linear function in (\ref{eq3_9}) is applied to the  relation in (\ref{eq2_13}), we obtain
\begin{align}
a_*=\sqrt{\dfrac{3 A_*^4 (g-1)^2-8 A_*^3 (g-1) g+6 A_*^2 g^2-g^2-2 g+3}{6}},\\
b_*=\sqrt{\dfrac{3 B_*^4 (g-1)^2-8 B_*^3 (g-1) g+6 B_*^2 g^2-g^2-2 g+3}{6}},
\end{align}
or
\begin{align}
a_*=\sqrt{\frac{A_*^4 \left(g^2-6 g+11\right)-6 A_*^2 \left(g^2-4 g+3\right)+8 A_*\left(g^2-3 
   g+2\right)-3 (g-1)^2}{6 A_*^2}},\\
b_*=\sqrt{\frac{\left(B_*^4-1\right) (g-1)^2}{2 A_*^2}-\frac{4 \left(B_*^3-1\right) \left(g^2-3
   g+2\right)}{3A_*}+B_*^2 (g-2)^2-g^2+4 g-3},
\end{align}
if (\ref{eq3_10}) is adopted.

It should be pointed out that, surface wrinkling of sinusoidal profiles may be incurred in both planar and curved film-substrate structures under appropriate stimuli such as axial compression \cite{38,39}, growth \cite{40,41,42,43}, or swelling \cite{44,45}, if the film is stiffer than the substrate, or equivalently, the $\beta$ should exceed a critical value. Furthermore, for planar structures, this critical value is proven to be around $1.74$ \cite{38,46} while for curved systems it is dependent on the geometrical parameters ranging from 1.18 to 1.6 \cite{21}. In this study, we are also concerned with such a typical instability, so only the case that the shear modulus of the inner layer is twice greater than that of the outer layer, i.e. $\beta>2$, is considered.
 
\begin{figure}[!h]
	\centering\includegraphics[scale=0.8]{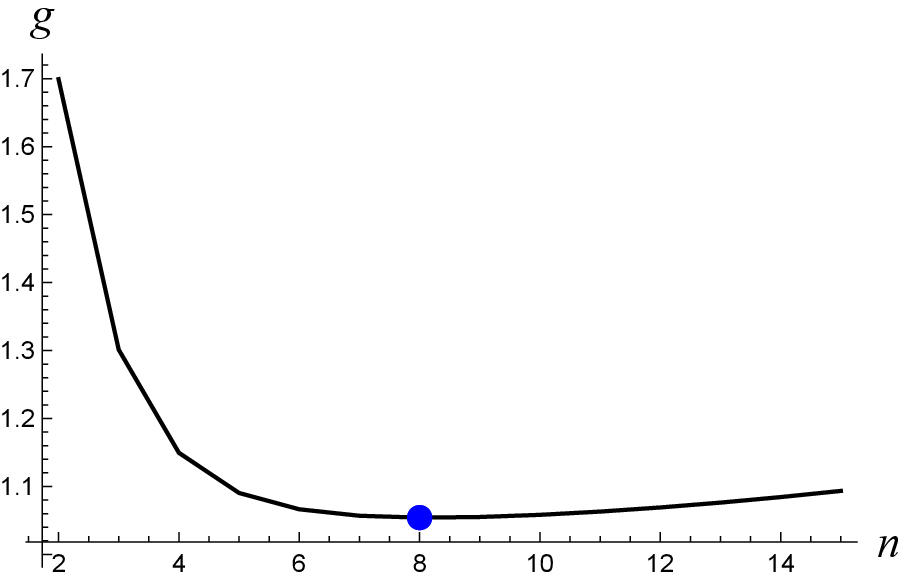}
	\caption{(Color online) Bifurcation curve for $A_*=0.67$, $B_*=0.7$,  and $\beta=100$ when (\ref{eq3_9}) is used.} The blue point highlights the minimum that corresponds to the critical growth factor $g_{cr}=1.05414$ associated with the critical wavenumber $n_{cr}=8$.\label{fig2}
\end{figure} 
 
It is pointed out earlier that the onset of surface wrinkling can be determined using determinant method once the growth gradient (\ref{eq3_9}) or (\ref{eq3_10}) is applied. The numerically obtained bifurcation condition can be viewed as a function of the dimensionless geometrical parameters $A_*$ and $B_*$, the ratio of the shear modulus $\beta$, the wavenumber $n$, and the growth factor $g$. In Figure \ref{fig2}, we sketch the dependence on the wavenumber $n$ of the growth factor $g$ for GGI by setting $A_*=0.67$, $B_*=0.7$, and $\beta=100$. It can be seen that this curve has a $U-$shape where the vertical coordinate of the minimum identifies the lowest growth factor triggering a sinusoidal pattern on the surface and the horizontal counterpart counts the waves. Consequently, we denote them as the critical growth factor $g_{cr}$ and the critical wavenumber $n_{cr}$, respectively. Likewise, we define the corresponding pattern as the critical pattern.

Next, we investigate the effect of the linear growth gradient on the wrinkling initiation and the corresponding critical pattern. For that purpose, the homogeneous growth case where both layers grow simultaneously in the same growth factor is taken into consideration for comparison. In other words, the growth gradient $f(R)$ (or $\widehat{f}(R)$) is identical to a constant in the parallel homogenous case. Furthermore, we employ another parameter $a_{*cr}$ instead of the critical growth factor since the former is more clear to depict how thick the structure grows prior to surface wrinkling. On the other hand, it is known that the critical wavenumber is mainly controlled by the circumference of the inner surface. As a result, the new parameter $a_{*cr}$ is more appropriate to depict the difference in critical state between different growth types. In addition to the growth type, all other parameters remain the same for these models, including two graded growth types and one homogeneous growth type.

\begin{figure}[!h]
	\centering
	\subfigure{\includegraphics[scale=0.7]{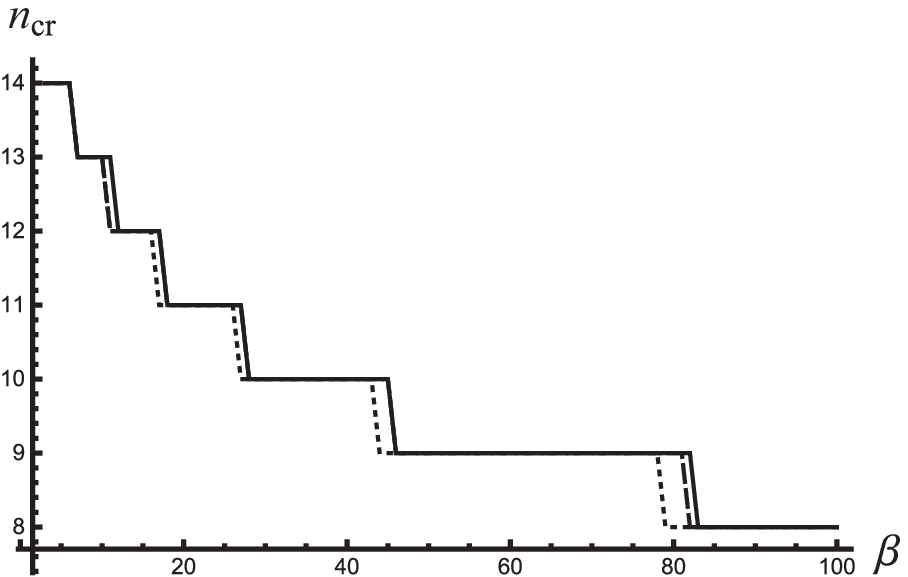}}
	\subfigure{\includegraphics[scale=0.7]{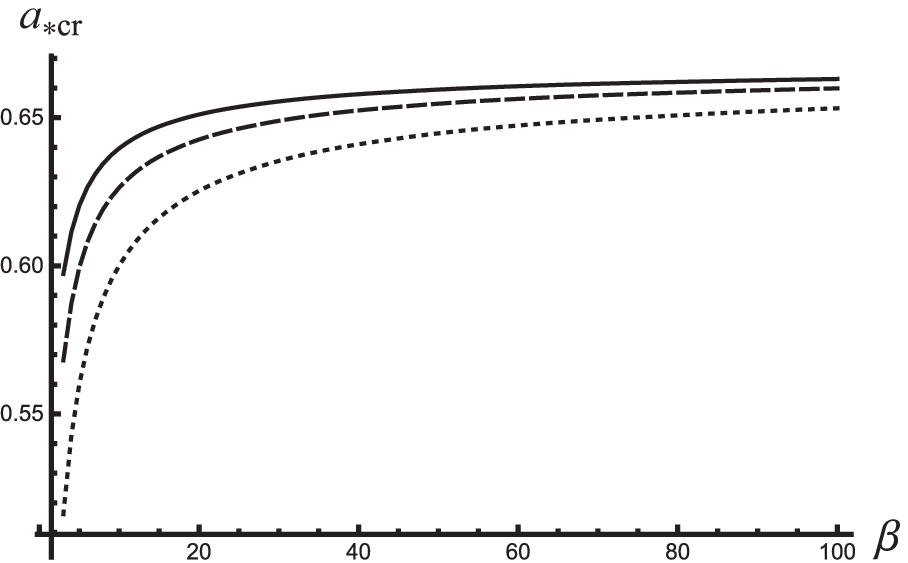}}
	\caption{Dependences of $n_{cr}$ and $a_{*cr}$ on $\beta$ when $A_*=0.67$ and $B_*=0.7$. The dashed lines denote the results for the homogeneous growth. The solids and dotted lines denote the counterparts for the GGI and GGII, respectively.}\label{fig3}
\end{figure}
It is observed that three free parameters are involved in the bifurcation condition, including the dimensionless inner and interfacial radii $A_*$ and $B_*$ and the ratio of the shear modulus $\beta$. As long as two of them are fixed, one could unravel the influence of the left parameter on the critical pattern, including the critical radius $a_{*cr}$ and the critical wavenumber $n_{cr}$. First, we let $A_*=0.67$ and $B_*=0.7$ and vary the ratio of the shear modulus $\beta$. Figure \ref{fig3} exhibits the results for all growth types. As expected, the critical wavenumber $n_{cr}$ is a monotonically decreasing function in $\beta$ for all growth types. In particular, the three curves are almost identical. For the critical radius $a_{*cr}$, the differences among these curves are minor and the solid curve is a little bit higher than the dashed one while the dotted line is a little bit lower than the dashed one. In other words, the bilayered structure grows less when it suffers a graded growth decaying from the inner layer for a specified $\beta$, leading to a higher $a_{*cr}$.  For GGII, the bilayered structure can grow thicker before surface wrinkles emerge. Nevertheless, such a discrepancy has almost no influence on the critical pattern.

\begin{figure}[!h]
	\centering
	\subfigure	{
			\includegraphics[scale=0.7]{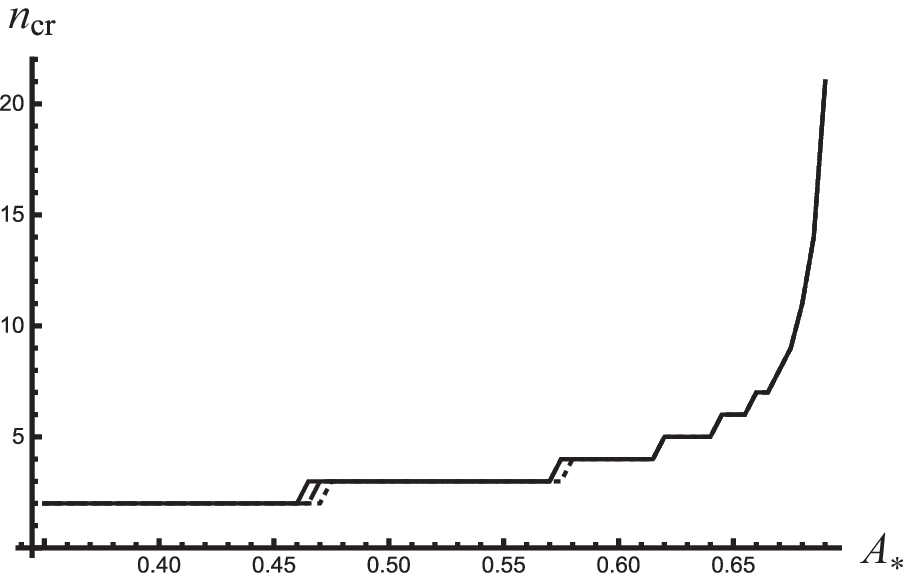}

	}
	\subfigure
	{
			\includegraphics[scale=0.7]{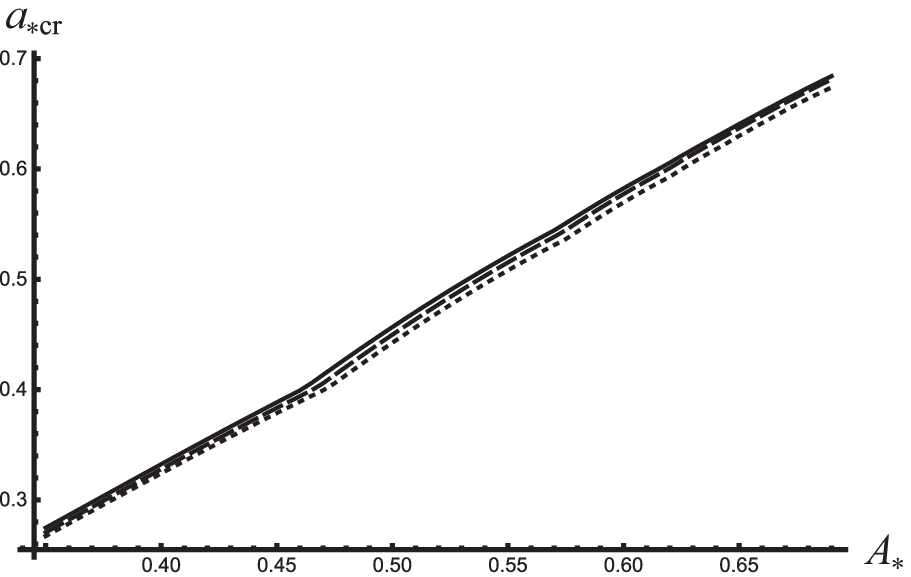}
	}
	\caption{Dependences of $n_{cr}$ and $a_{*cr}$ on $A_*$ when $B_*=0.7$ and $\beta=100$. The dashed lines denote the results for the homogeneous growth. The solids and dotted lines denote the counterparts for the GGI and GGII, respectively.}\label{fig4}
\end{figure}

\begin{figure}[!h]
	\centering
	\subfigure
	{\includegraphics[scale=0.72]{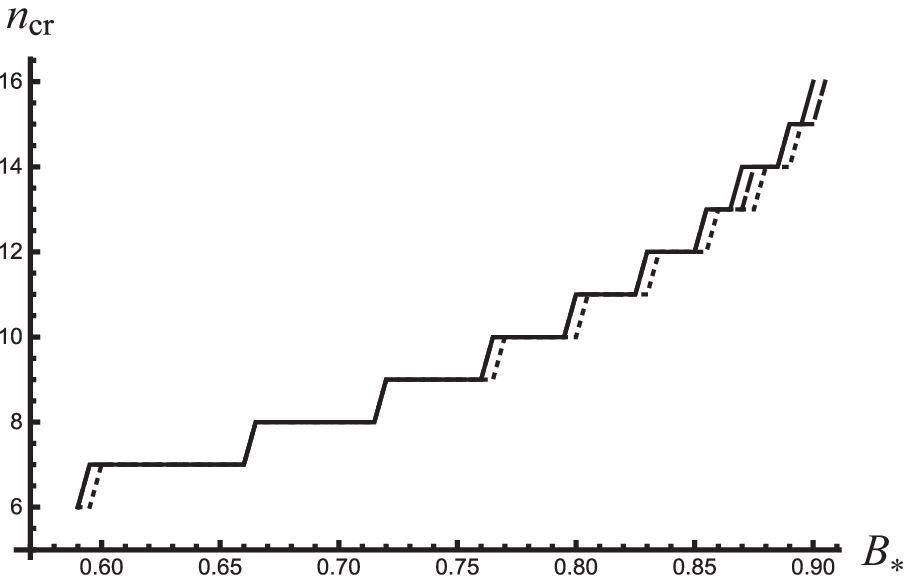}}
	\subfigure
	{\includegraphics[scale=0.72]{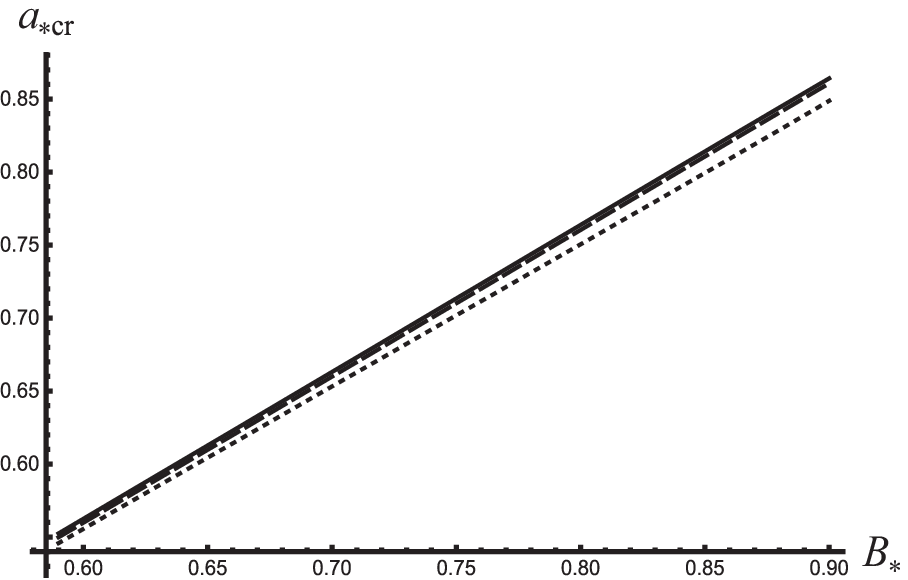}}
	\caption{Dependences of $n_{cr}$ and $a_{*cr}$ on $B_*$ when $h_{*}=B_*-A_*=0.03$ and $\beta=100$ for the graded and homogeneous growth types.}\label{fig5}
\end{figure}

Then we explore the effect of the dimensionless inner radius  $A_*$ by specifying $\beta=100$ and $B_*=0.7$ and plot the results in Figure \ref{fig4}. The solid lines, dashed lines, and dotted lines (see also Figure \ref{fig5}) remain the same representations as those in Figure \ref{fig3}. Remarkably, the three curves for the critical wavenumber $n_{cr}$ almost overlap each other. It is emphasized that the dependence of $n_{cr}$ on $A_*$ is highly nonlinear and the critical wavenumber experiences a fast increase as $A_*$ is close to $B_*$. Bearing in mind that the critical wavenumber is of $O(1/(\beta^{\frac{1}{3}} h_*))$ ($h_*=B_*-A_*$) when the inner layer grows solely \cite{20}, so the tendencies in Figures \ref{fig3} and \ref{fig4} are also well captured by this scaling law. Moreover, it is found that the $a_{*cr}$ is practically linear as a function of the inner radius $A_*$ for the three growth types, and the deviations among these curves are negligible. This again confirms that the linear growth gradient has a weak influence on the critical pattern.

Finally, we focus on the situation where the interfacial radius $B_*$ is varied. In this case, the dimensionless thickness of the inner layer $h_*$, which has been defined earlier, is assigned by 0.03 and the modulus ratio is given by $\beta=100$. Consequently, changing $B_*$ is equivalent to altering the thickness of the outer layer with the thickness of the inner layer being fixed. The dependences of $n_{cr}$ and $a_{*cr}$ on $B_*$ are displayed in Figure \ref{fig5}. Similar to the results in Figure \ref{fig4}, these three growth types generate qualitatively and quantitively analogical predictions. In detail, the tendencies for $n_{cr}$ and $a_{*cr}$ when $B_*$ is increased remain the same as that in Figure \ref{fig4}. Note that in this case the inner radius $A_*$ is also changed as $B_*$ varied, so this may naturally cause the similar results as we observed. On the other hand, we can fix the thickness of the whole structure and change the interfacial radius. Correspondingly, the thicknesses for the inner and outer layers will be modified at the same time. However, the main aim is to clarify the influence of the growth gradient on the critical pattern, and we have shown three case studies. Specifically, each case presents the same conclusion that the growth gradient plays a minor role in regulating the critical pattern. As a result, we stop giving additional examples. 

In conclusion, we have conducted a parametric study to examine the role of a linear growth gradient in determining the surface wrinkling. It is worth mentioning that we actually reproduce all procedures in obtaining Figures \ref{fig3}--\ref{fig5} when a constant growth drives the inner layer and there is no growth in the outer layer. It turns out that there is also no essential difference compared to the given curves. For the sake of brevity, we omit these curves and directly narrate the results here.

\section{Finite element validations}
In this section, we aim at validating the theoretical predictions based on the incremental theory by employing finite element analysis (FEA) in commercial software Abaqus \cite{47}. To simulate the growth process in Abaqus, we refer to \cite{48} and adopt an analogy to thermal expansion. In addition, we apply a graded temperature field that is in accordance with (\ref{eq3_9}) to model the graded growth. As a result, the physical model for FEA is well-established. 

In the previous section, two specific graded growth conditions, labelled by GGI and GGII, respectively, are studied. It is found that both GGI and GGII provide similar theoretical predictions. Consequently, we only establish a FE model for GGI and then carry out a FEA for a validation.

In our simulations, both the four-node plane-strain element (CPE4R) and the eight-node plane-strain element (CPE8R) are used. It is worth mentioning that the former mesh performs well when the parameter $\beta$ is lower ($\beta<10$) while the latter is better for greater $\beta$. In particular, the General Static module is employed. In doing so, not only the critical buckled state but also the pattern evolutions can be identified. Meanwhile, the bifurcation threshold can be determined following the same methodology in Liu et al. \cite{29} and we omit details for brevity. Bearing in mind that the main task in this section is to validate the theoretical results, we only exhibit the results for the primary bifurcation state in the subsequent analysis.  

\begin{figure}[!h]
	\centering
	{\subfigure{\includegraphics[scale=0.72]{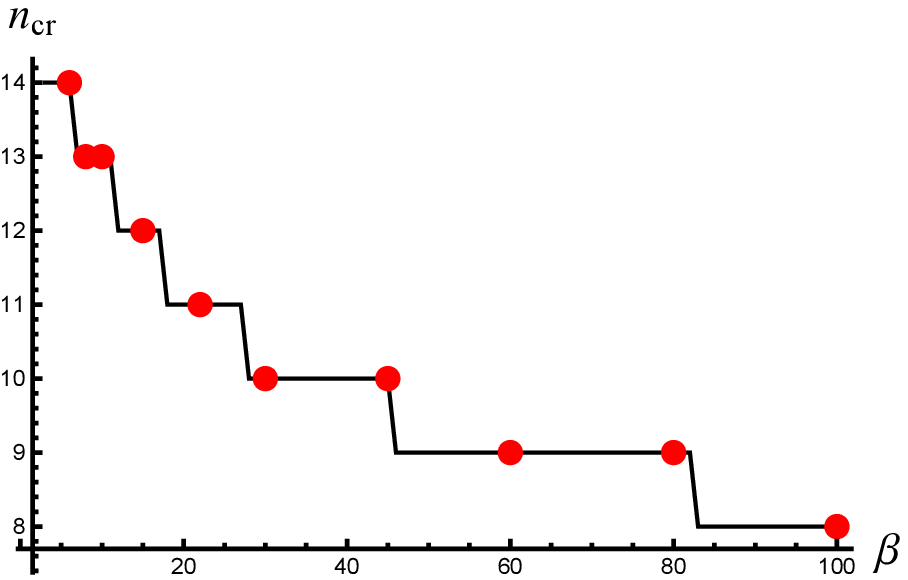}}
%		\hspace{2mm}
	\subfigure{\includegraphics[scale=0.72]{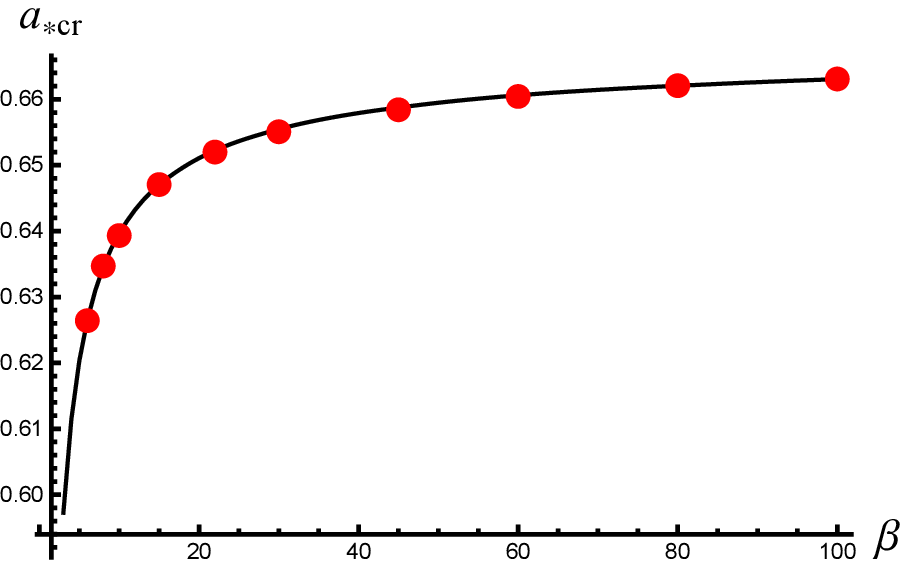}}
	\caption{(Color online) Dependences of $n_{cr}$ and $a_{*cr}$ on $\beta$ when $A_*=0.67$ and $B_*=0.7$. The solid lines express the theoretical results while the red dots denote the counterparts in FEA.}
		\label{fig6}}
\end{figure}

\begin{figure}[!h]
	\centering
	{\subfigure{\includegraphics[scale=0.72]{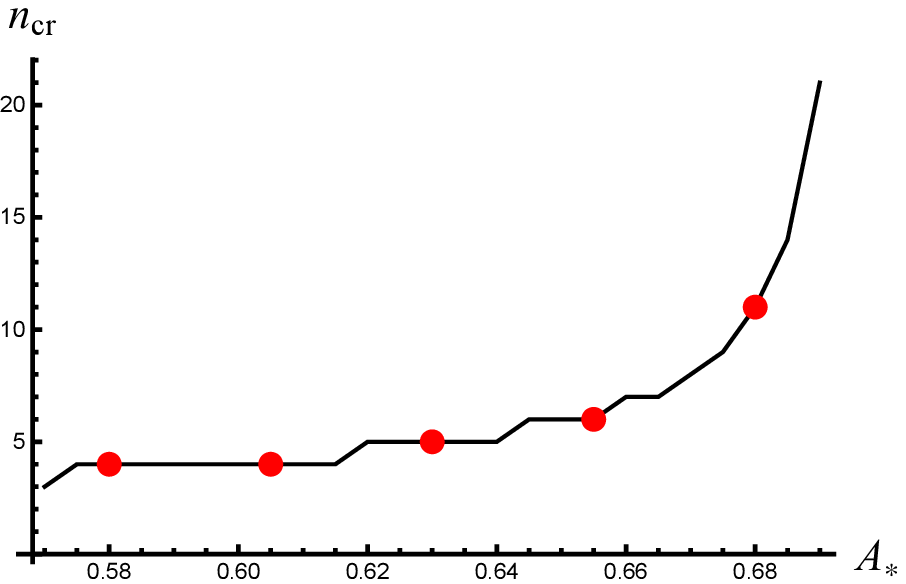}}
		%		\hspace{2mm}
		\subfigure{\includegraphics[scale=0.72]{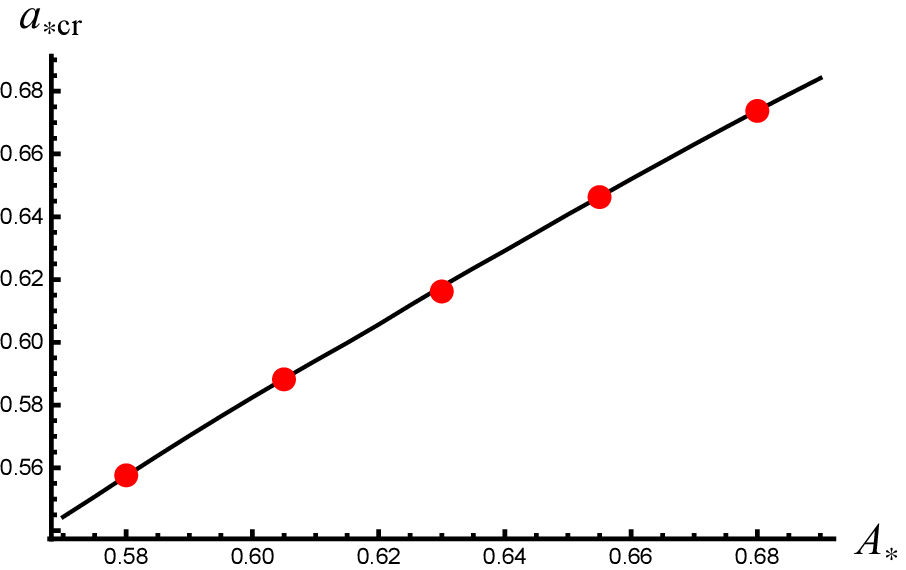}}
		\caption{(Color online) Dependences of $n_{cr}$ and $a_{*cr}$ on $A_*$ when $B_*=0.7$ and $\beta=100$ for the theoretical and numerical solutions.}
		\label{fig7}}
\end{figure}

\begin{figure}[!h]
	\centering
	{\subfigure{\includegraphics[scale=0.72]{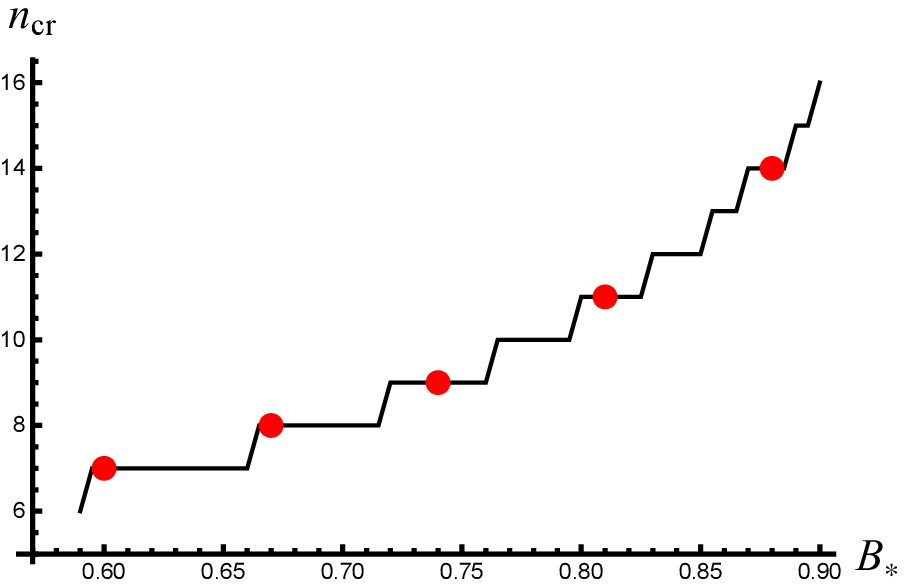}}
		%		\hspace{2mm}
		\subfigure{\includegraphics[scale=0.72]{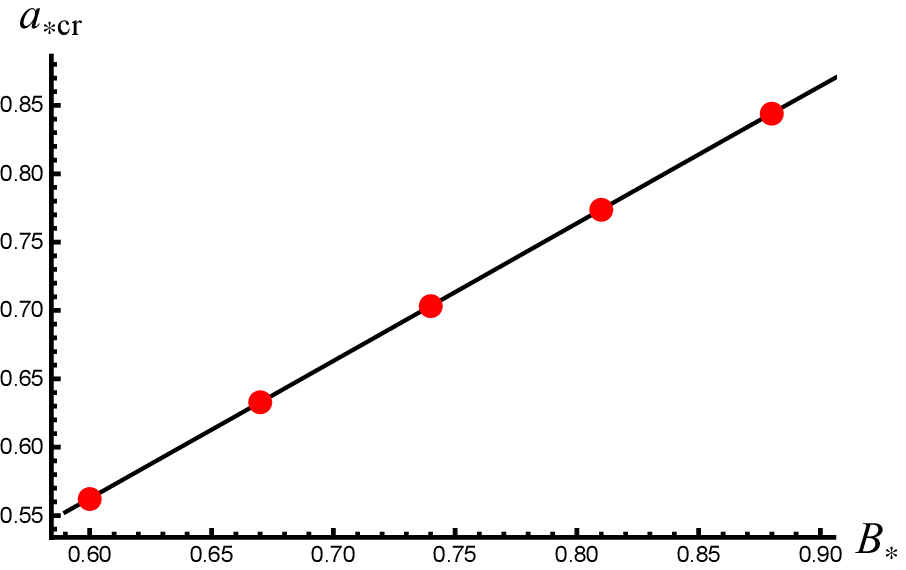}}
		\caption{(Color online) Dependences of $n_{cr}$ and $a_{*cr}$ on $B_*$ when $h_*=B_*-A_*=0.03$, and, $\beta=100$ for the theoretical and numerical solutions.}
		\label{fig8}}
\end{figure}

At first, we study the case where the ratio of the shear modulus $\beta$ is varied. Figure \ref{fig6} depicts the critical wavenumber $n_{cr}$ and the critical radius $a_{*cr}$ as functions of $\beta$ when other parameters are fixed by $A_*=0.67$ and $B_*=0.7$. The results based on the theoretical model are denoted by solid lines while those from FEA are denoted by red dots (see also Figures \ref{fig7} and \ref{fig8}). It can be seen that the theoretical predictions agree extremely well with FE solutions, which indicates the validity of both the FE model and the theoretical model.

In Figure \ref{fig7}, the normalized inner radius $A_*$ becomes the free parameter while $\beta$ is set to be $100$ instead. As the normalized interfacial radius $B_*$ is fixed to be 0.7, varying $A_*$ is equivalent to altering the thickness of the inner layer. Similarly, both theoretical and FE models offer the same prediction. Yet the critical wavenumber $n_{cr}$ and critical {color{blue}radius} $a_{*cr}$ are monotonically increasing functions with respect to $A_*$, which manifests that a bilayered tubular tissue with thinner inner layer produces a lower wavelength. 

Figure \ref{fig8} shows the critical wavenumber $n_{cr}$ and the critical radius $a_{*cr}$ versus $B_*$. As we expected, both solutions agree well. It is found that all curves in Figures \ref{fig7} and \ref{fig8} maintain the same profiles such that similar conclusions from Figure \ref{fig7} can also be drawn in Figure \ref{fig8}. It is found that the modulus ratio $\beta$ develops into the main parameter in determining $a_{*cr}$ instead of $A_*$ and $B_*$. 

Currently, the theoretical model and the finite element counterpart have been mutually validated. In the next section, the FE model will be further used to elucidate the role played by the growth gradient in shaping the surface morphology.

\section{Post-buckling evolution}

In the previous sections, we have revealed the influence of a linear growth function on the critical pattern and have established a robust FE model in Abaqus to simulate graded growth. It turns out that the bifurcation nature of growing tubular tissues is mainly dominated by the geometrical and material parameters, not the growth gradient. In our previous study in \cite{29}, it is known that different pattern transitions can be induced by a homogeneous growth in graded tubular tissues. This implies that material gradient can alter post-buckling evolution. In this section, we are left to shed light on whether a growth gradient can have same effect. To this end, we adopt the FE model developed in Section 4 and track the post-buckling evolution by carrying out a fully nonlinear analysis. From the linear bifurcation analysis, it is known that either a homogeneous growth factor or a graded one can create a wavy pattern of sinusoidal wrinkles in growing tubular tissues. Therefore, we focus on how a wavy pattern evolves driven by a graded growth. As outlined earlier, only the post-buckling behaviors of GGI will be shown later.

Before proceeding further, we recall some necessary results for the secondary bifurcation-induced pattern transitions in film-substrate structures. For planar film-substrate structures and according to an asymptotic analysis, Fu and Cai \cite{49} analytically derived a necessary condition $\beta>5.8$ that period-doubling pattern may exist. By means of FEM and Abaqus, more complicated patterns such as ridge can be studied numerically and a phase diagram for different patterns can be drawn \cite{50}. It turns out that the ratio of the shear modulus $\beta$ is critical in determining the pattern transition. For the bilayered tubular tissues studied in the present paper, Li et al. \cite{14} showed two typical post-buckling evolutions caused either by inner layer growth or outer layer growth when $\beta$ is large. This indicates that different growth types may incur various patterns.

\begin{figure}
\centering\includegraphics[scale=0.5]{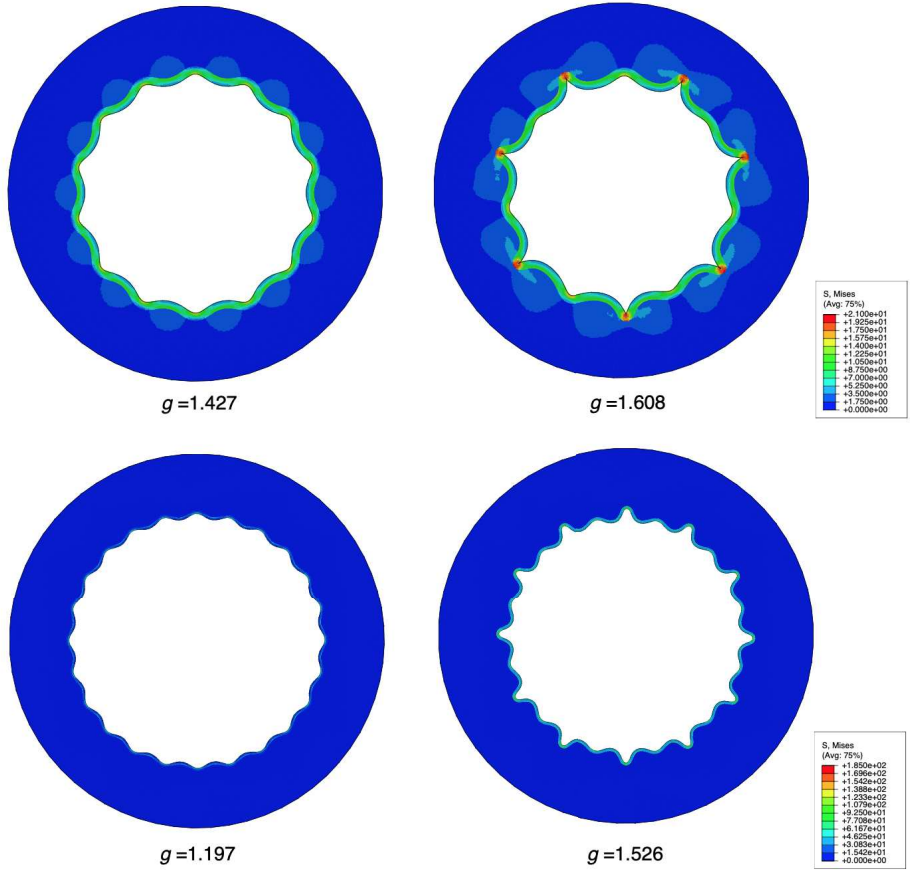}
\caption{(Color online) Pattern evolutions of a bilayered tubular tissue suffering the graded growth described by (\ref{eq3_9}). The top line represents the results for $\beta=6$, $A_*=0.67$, $B_*=0.7$ while the bottom line corresponds to $\beta=60$, $A_*=0.69$, $B_*=0.7$. The value below each subfigure depicts the growth factor $g$ in equation (\ref{eq3_9}) at that state.}\label{fig9}
\end{figure}

We point out that in a structure of infinite length, period-doubling profile becomes possible only if $\beta$ exceeds a critical value $5.8$. For a tubular structure, the critical wavenumber can either be an even number or an odd one. If we still expect to observe a perfect period-doubling pattern, the associated wrinkling mode should possess an even number. We then pay attention to two parametric settings, i.e., $\beta=6$, $A_*=0.67$, $B_*=0.7$ and $\beta=60$, $A_*=0.69$, $B_*=0.7$. In doing so, the critical wavenumbers  read $n_{cr}=14$ and $n_{cr}=22$, respectively. The post-buckling evolutions are sketched in Figure \ref{fig9}. Seen from the top figures in Figure \ref{fig9}, the sinusoidal pattern develops into a period-doubling morphology where every period contains a crease and a wrinkle for $\beta=6$. Since there exists a sharp tip, we call it the crease formation \cite{40}. Yet such a special pattern will give way to the ordinary period-doubling mode when $\beta=60$. We mention that such a special mode has been experimentally observed in graded structures \cite{51}. Meanwhile, our previous study in graded growing tubular tissues also presented this mode \cite{29} when the shear modulus decays exponentially. 

\begin{figure}[!h]
	\centering\includegraphics[scale=0.5]{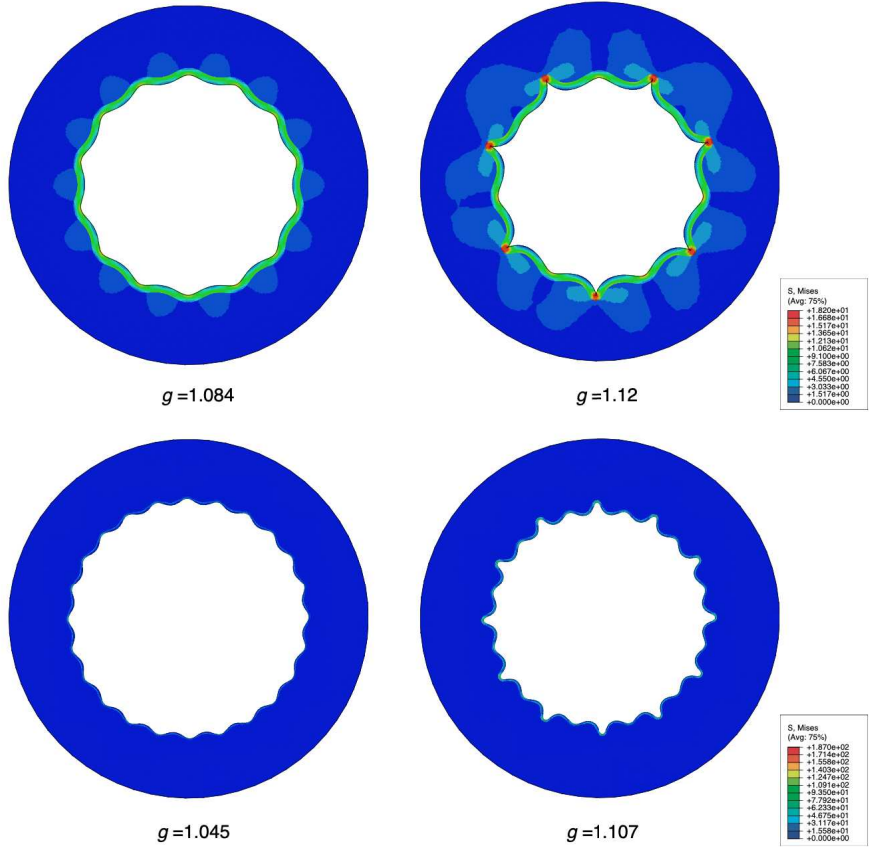}
	\caption{(Color online) Pattern evolutions of a bilayered tubular tissue for the homogeneous growth type where both layers grow simultaneously. The top line represents the results for $\beta=6$, $A_*=0.67$, $B_*=0.7$ while the bottom line corresponds to $\beta=60$, $A_*=0.69$, $B_*=0.7$. The value below each subfigure depicts associated the growth factor.}\label{fig10}
\end{figure}

To reveal the influence of the linear growth gradient on the pattern transitions in tubular tissues, we present the post-buckling evolutions for the homogeneous growth type in Figure \ref{fig10}. Note that the all parameters remain the same as those in Figure \ref{fig9}. Similar to the comparisons for the critical pattern, we also find that the two growth types generate the same prediction for the pattern transition. In particular, it is the ratio of the shear modulus that can make a major difference.

\begin{figure}[!h]
	\centering
	\subfigure[The grade growth type GGI.]
        {\includegraphics[scale=0.7]{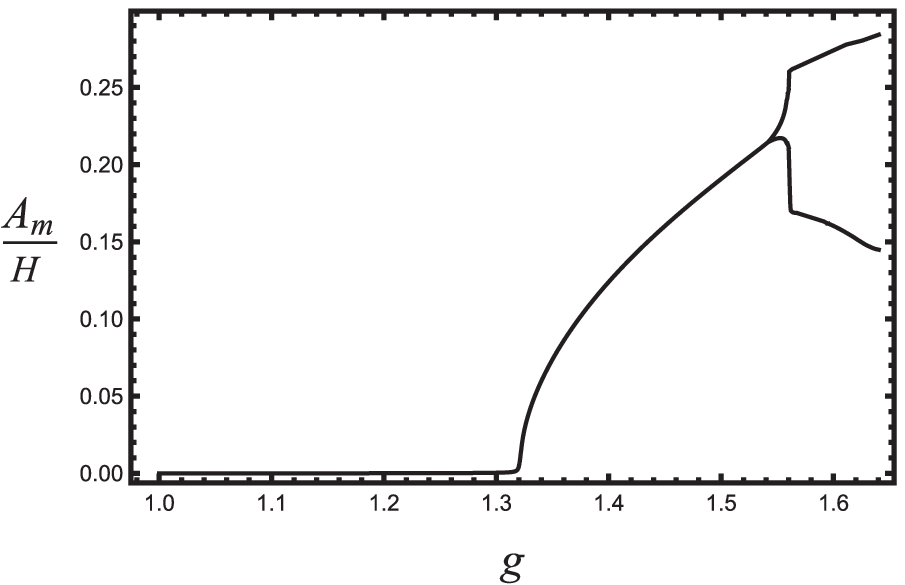}{\label{fig11a}}}
	\hspace{2mm}
	\subfigure[The homogeneous growth type.]
	{\includegraphics[scale=0.7]{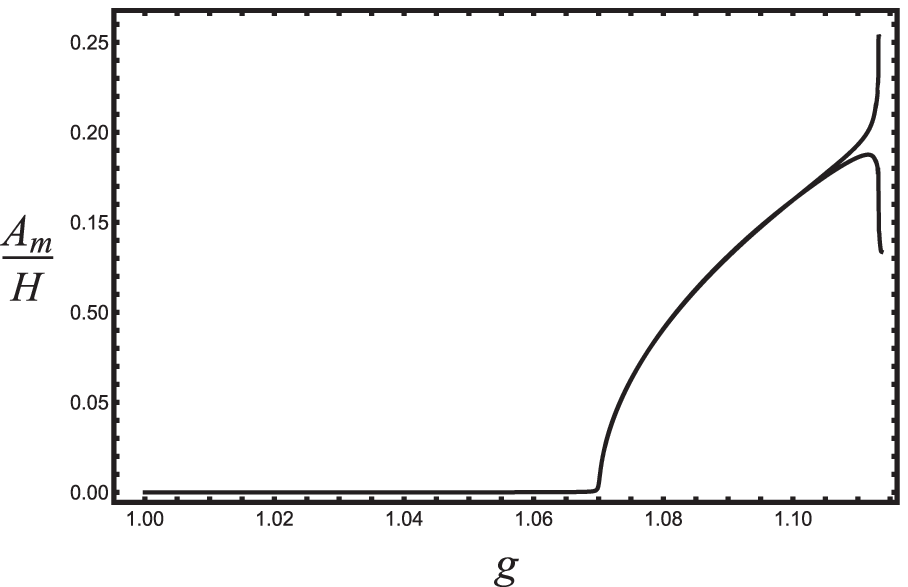}{\label{fig11b}}}
\caption{Bifurcation diagrams for the period-doubling pattern containing self-contact. The parameters are given by  $\beta=6$, $A_*=0.67$, and $B_*=0.7$.}
	\label{fig11}
\end{figure}

Finally, we plot the bifurcation diagrams of the period-doubling morphology with creases based on FEA in Figure \ref{fig11} for both growth types. The counterparts for the ordinary period-doubling mode are depicted in Figure \ref{fig12}. In the homogeneous case, the growth function (\ref{eq3_9}) is reduced to a constant growth factor $g$. The vertical axis denotes the normalized amplitude $A_m/H$ where $A_m$ is defined by computing the distance between a peak and a valley and $H=C-A$ gives the thickness of the structure. The horizontal axis stands for the growth factor $g$ that drives the growth process. It can be seen from Figures \ref{fig11} that the structure bifurcates at $g_{cr}=1.321$ and $g_{cr}=1.0695$, respectively, resulting in a nontrivial amplitude. Then the surface pattern evolves with an increasing amplitude until the growth factor $g$ exceeds another critical value $g_{se}=1.555$ in Figure (\ref{fig11a}) and $g_{se}=1.105$ in Figure (\ref{fig11b}), a period-doubling pattern occurs. In addition, both growth types generate the period-doubling formation when the dimensionless amplitude $A_m/H$ is around $0.2$. Seen from Figure \ref{fig12}, surface wrinkling can be induced when $g_{cr}=1.073$ and $g_{cr}=1.016$, respectively, and a secondary bifurcation leading to period-doubling morphology is triggered when the growth factor reaches $g_{se}=1.449$ in Figure \ref{fig12a} and $g_{se}=1.095$ in Figure \ref{fig12b}, respectively. Furthermore, the ordinary period-doubling pattern starts when $A_m/H$ reaches near 0.15.

{\color{blue}
\begin{figure}[!h]
	\centering
	\subfigure[The grade growth type GGI.]
        {\includegraphics[scale=0.7]{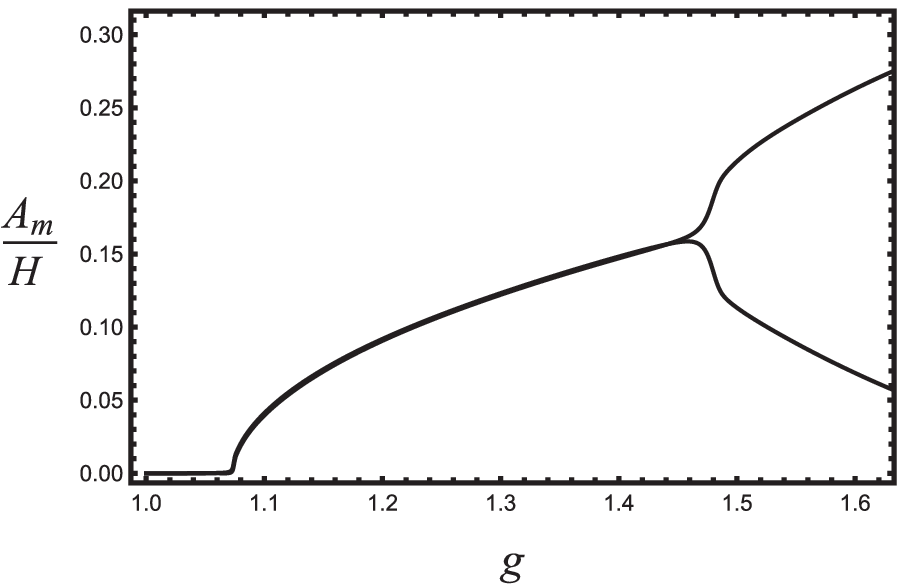}{\label{fig12a}}}
	\hspace{2mm}
	\subfigure[The homogeneous growth type.]
	{\includegraphics[scale=0.7]{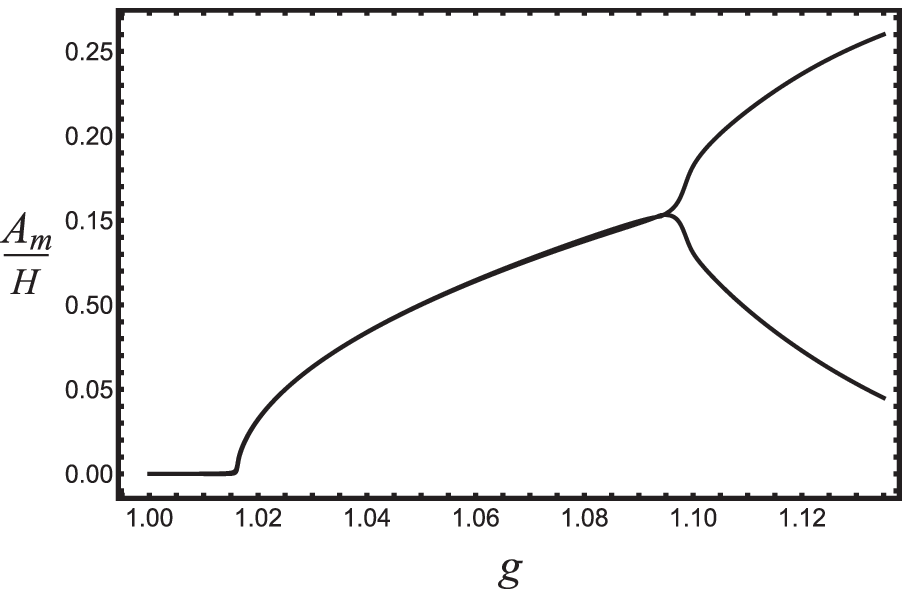}{\label{fig12b}}}
\caption{Bifurcation diagrams for the ordinary period-doubling pattern. The parameters are given by  $\beta=60$, $A_*=0.69$, and $B_*=0.7$.}
	\label{fig12}
\end{figure}}

\section{Conclusion}
Under the plane-strain setting, the influence of an inhomogeneous growth field on surface instabilities and pattern evolutions in tubular bilayered tissues was addressed in this paper. Within the framework of finite elasticity and by use of the incremental theory, we theoretically characterized the deformation prior to surface wrinkling and identified the critical condition for surface wrinkling for a general material model and a general growth distribution. For illustrative examples, we employed the neo-Hookean model and specified a growth field that decays linearly either from the inner surface or from the outer surface. The bifurcation curves were illustrated by numerically solving the bifurcation condition using determinant method. A detailed comparison between graded growth and homogeneous growth types was performed. It is found that the prescribed growth gradient has a minor influence on the critical radius and the critical wavenumber. Furthermore, a finite element model, validated by the theoretical solution, was established by an analogue to thermal expansion referring to \cite{48}. According to the FE model, a post-buckling analysis was carried out to reveal the pattern transitions. Since previous analysis reveals that the ratio of the shear modulus $\beta$ is a critical parameter in determining the pattern evolution \cite{49,50}, we also focused on two situations, i.e. $\beta=6$ and $\beta=60$ and exhibited the corresponding patterns for both growth types, respectively. It turns out that the initial bifurcation generates a wrinkled surface pattern and the wrinkled mode will evolve into a period-doubling mode where each period includes a crease and a wrinkle when the modulus ratio is given by $\beta=6$. However, this special mode will give way to a normal period-doubling mode, as described by Fu and Cai \cite{49}, when $\beta=60$. Moreover, we plotted the bifurcation diagrams for both patterns. Finally, the corresponding pattern evolutions for the homogeneous growth type were shown. In accordance with our expectation, the corresponding results remain qualitatively the same as that for the graded growth. In conclusion, our analysis reveals that assuming a homogeneous growth field does provide a realistic approximation in tubular tissues with fixed outer surface and can capture the main features of growth-induced surface instabilities. Also, it gives further evidence to the viewpoint that the thickness parameter dominates the critical wavenumber as well as the wavy pattern, but the morphology incurred by a secondary bifurcation is mainly controlled by the ratio of the shear modulus. 

\section*{Acknowledgments}
The work was supported by National Natural Science Foundation of China (Project Nos: 12072227 and 12021002). The Abaqus simulations were carried out on TianHe-1 (A) at the National Supercomputer Center in Tianjin, China.

\begin{appendix}
\section*{Appendix A. The explicit incremental equations for the inner layer and the associated boundary conditions and continuity conditions}
We write explicitly the incremental equations (\ref{eq2_18}) associated with the boundary conditions in (\ref{eq2_21}) and the continuity conditions in (\ref{eq2_22}) as follows:
\begin{align}
	p^{*}_{,r}=&~(\mathcal{A}_{1111}+p)u_{,rr}+\dfrac{1}{r^2}\mathcal{A}_{2121}u_{,\theta \theta}+\dfrac{1}{r}(\mathcal{A}_{1111}+\mathcal{A}_{1122}-\mathcal{A}_{2211}+p+r\mathcal{A}_{1111}'+rp')u_{,r}\notag\\
	&-\dfrac{1}{r^2}(\mathcal{A}_{2222}+p-r\mathcal{A}_{2222}')u+\dfrac{1}{r}(\mathcal{A}_{1122}+\mathcal{A}_{2112}+p)v_{,r \theta}\notag\\
	&-\dfrac{1}{r^2}(\mathcal{A}_{2121}+\mathcal{A}_{2222}+p-r\mathcal{A}_{1122}')v_{,\theta},
	\label{eq2_23}
\end{align}

\begin{align}
	\dfrac{1}{r}p^{*}_{,\theta}=&~\mathcal{A}_{1212}v_{,rr}+\dfrac{1}{r^2}(\mathcal{A}_{2222}+p)v_{,\theta \theta}+\dfrac{1}{r}(\mathcal{A}_{1212}-\mathcal{A}_{1221}+\mathcal{A}_{2112}+r\mathcal{A}_{1212}')v_{,r}\notag\\&-\dfrac{1}{r^2}(\mathcal{A}_{2121}+r\mathcal{A}_{1221}'+rp')v+\dfrac{1}{r}(\mathcal{A}_{1221}+\mathcal{A}_{2211}+p)u_{,r \theta}\notag\\&+\dfrac{1}{r^2}(\mathcal{A}_{2121}+\mathcal{A}_{2222}+p+r\mathcal{A}_{1221}'+rp')u_{,\theta},
	\label{eq2_24}
\end{align}

\begin{align}
	&(\mathcal{A}_{1111}+p-\mathcal{A}_{1122})u_{,r}-p^{*}=0,~~v_{,r}+\dfrac{1}{r}(u_{,\theta}-v)=0,~~\textrm{on}~~r=a,\notag\\
	&\bar{u}=0,~~\bar{v}=0,~~\textrm{on}~~r=C,
\end{align}
\begin{align}
	&(\mathcal{A}_{1111}+p-\mathcal{A}_{1122})u_{,r}-p^{*}=(\bar{\mathcal{A}}_{1111}+\bar{p}-\bar{\mathcal{A}}_{1122})\bar{u}_{,r}-\bar{p}^{*},~~\textrm{on}~~r=b,\notag\\
	&\mathcal{A}_{1212}\left(v_{,r}+\dfrac{1}{r}(u_{,\theta}-v)\right)=\bar{\mathcal{A}}_{1212}\left(\bar{v}_{,r}+\dfrac{1}{r}(\bar{u}_{,\theta}-\bar{v})\right),~~\textrm{on}~~r=b,\notag\\
	&u=\bar{u},~~v=\bar{v},~~~~\textrm{on}~~r=b.
\end{align}
It is worth mentioning that the associated equations for the outer layer can be obtained by suitable changes of notation to equations (\ref{eq2_23}) and (\ref{eq2_24}).

\section*{Appendix B. The components of coefficient matrices in (\ref{eq3_2})--(\ref{eq3_4})}
\begin{align*}
r^4\mathcal{A}_{1212}m_{41}&=(n^2-1)((1-n^2)\mathcal{A}_{2121}-r\mathcal{A}_{2121}'-r^2\mathcal{A}_{1221}''-r^2p'')\\
r^3\mathcal{A}_{1212}m_{42}&=\mathcal{A}_{2121}+r^2(-2\mathcal{A}_{1212}''+\mathcal{A}_{1221}''+p'')\\&~~~~+r(-4\mathcal{A}_{1212}'+4\mathcal{A}_{1221}'-2\mathcal{A}_{2112}'+\mathcal{A}_{2121}'+2p')\\&~~~~+n^2(\mathcal{A}_{1111}-2\mathcal{A}_{1122}-2\mathcal{A}_{2112}+\mathcal{A}_{2222}+r(\mathcal{A}_{1111}'-2\mathcal{A}_{1122}'-2\mathcal{A}_{1221}'+\mathcal{A}_{2222}')),\\
r^2\mathcal{A}_{1212}m_{43}&=3\mathcal{A}_{1221}-6\mathcal{A}_{1212}-3\mathcal{A}_{2112}+\mathcal{A}_{2121}-r^2\mathcal{A}_{1212}''\\&~~~~+n^2(\mathcal{A}_{1111}-2\mathcal{A}_{1122}-\mathcal{A}_{1221}-\mathcal{A}_{2112}+\mathcal{A}_{2222})\\&~~~~+r(2\mathcal{A}_{1221}'-8\mathcal{A}_{1212}'-\mathcal{A}_{2112}'+p'),\\
r\mathcal{A}_{1212}m_{44}&=\mathcal{A}_{1221}-6\mathcal{A}_{1212}-\mathcal{A}_{2112}-2r\mathcal{A}_{1212}',\\
t_{21}&=\dfrac{1-n^2}{r}\left(\mathcal{A}_{2121}+r(\mathcal{A}_{1221}'+p')\right),\\
t_{22}&=\mathcal{A}_{1221}-2\mathcal{A}_{1212}-2\mathcal{A}_{2112}+\mathcal{A}_{2121}\\&~~~~+
n^2(\mathcal{A}_{1111}-2\mathcal{A}_{1122}-\mathcal{A}_{1221}+\mathcal{A}_{2222}+p)+r(\mathcal{A}_{1221}'-2\mathcal{A}_{1212}'+p''),\\
t_{23}&=r(4\mathcal{A}_{1212}-\mathcal{A}_{1221}+\mathcal{A}_{2112}+r\mathcal{A}_{1212}').
\end{align*}
\end{appendix}

%\noindent If maintaining .bib file for references, then please use "RS.bst" to generate the references.

%\noindent Example:

%\verb+\bibliographystyle{RS}+ %%%% .BST file

%\verb+\bibliography{sample}+ %%%%% .Bib file

\end{document}